\documentclass{aa}  
\usepackage{graphicx}
\usepackage{natbib}
\usepackage{pdfpages}
\usepackage{color}
\usepackage[varg]{txfonts}
\usepackage{amsmath}
\usepackage{subcaption}
\usepackage[normalem]{ulem}
\usepackage{txfonts}
\usepackage[breaklinks=true]{hyperref} 
\usepackage{orcidlink}
\usepackage[none]{hyphenat}

\def\apjl {ApJL}
\def\apjs {ApJS}

\def\aap {A\&A}

\begin{document}

\newcommand{\angstrom}{\textup{\AA}}

\definecolor{julian}{RGB}{0, 0, 204}
\newcommand{\jm}[1]{\textcolor{julian}{(J:)\ #1}}
\definecolor{Hernan}{RGB}{128, 0, 255}
\newcommand{\HM}[1]{\textcolor{Hernan}{HM:\ #1}}
\definecolor{vale}{rgb}{0.8,0,0}
\newcommand{\VC}[1]{\textcolor{vale}{(VC:) \ #1}}
\newcommand{\ANR}[1]{\textcolor{orange}{(ANR:) \ #1}}
\newcommand{\MA}[1]{\textcolor{magenta}{(chuti:)\ #1}}

\title{Reconstructing orbits of galaxies in extreme regions (ROGER). IV. Unveiling galaxy evolution patterns in OmegaWINGS clusters.}
\titlerunning{Unveiling galaxy evolution patterns }
 
\author{ Hern\'an Muriel\inst{1,2,3} \orcidlink{0000-0002-7305-9500}
         \and David Pérez-Millán\inst{4,5}\orcidlink{0000-0002-4507-9571}
          \and Mart\'in de los Rios\inst{6,7,8}\orcidlink{0000-0003-2190-2196}
          \and Andrea Biviano \inst{4,9}\orcidlink{0000-0002-0857-0732}
          \and Valeria Coenda\inst{1,2}\orcidlink{0000-0001-5262-3822}
          \and H\'ector J. Mart\'inez\inst{1,2}\orcidlink{0000-0003-0477-5412}
          \and Andr\'es N. Ruiz\inst{1,2}\orcidlink{0000-0001-5035-4913}
          \and  Benedetta Vulcani\inst{10}\orcidlink{0000-0003-0980-1499}
          \and Selene Levis\inst{1}\orcidlink{0000-0003-1887-776X}
}
   \institute{Instituto de Astronom\'ia Te\'orica y Experimental (IATE), CONICET - UNC, Laprida 854, X5000BGR, C\'ordoba, Argentina
             \and
             Observatorio Astron\'omico, Universidad Nacional de C\'ordoba, Laprida 854, X5000BGR, C\'ordoba, 
             Argentina
            \and
            INAF-Osservatorio Astronomico di Trieste, via G.B. Tiepolo, 11 - I-34143 Trieste, Italy
             \and 
             Instituto de Radioastronomia y Astrofisica, UNAM, Campus Morelia, Michoacán CP 58089, Mexico
             \and
            INAF–Osservatorio di Astrofisica e Scienza dello Spazio di Bologna, Via Gobetti 93/3, 40129 Bologna, Italy   
             \and
            Departamento de F\'isica Te\'orica, Universidad Aut\'onoma de Madrid, 28049 Madrid, Spain
            \and
            Instituto de F\'isica Te\'orica, IFT-UAM/CSIC, C/ Nicolás Cabrera 13-15, Universidad Aut\'onoma de Madrid, Cantoblanco, Madrid 28049, Spain
            \and
            SISSA -  International School for Advanced Studies, Via Bonomea 265, 34136 Trieste, Italy
            \and
            IFPU Institute for Fundamental Physics of the Universe, via Beirut, 2 - I-34014 Trieste, Italy
            \and  
            INAF- Osservatorio astronomico di Padova, Vicolo Osservatorio 5, I-35122 Padova, Italy
            }           

   \date{Received XXXX; accepted XXXX}
\authorrunning{H. Muriel}

\abstract
{Clusters of galaxies have proven to be efficient systems in modifying various properties of galaxies, such as star formation or morphology. However, projection effects impose serious challenges in determining how, when, and to what extent galaxies are affected by the cluster environment.}
{Using innovative techniques to classify galaxies based on their history within the cluster, we aim to determine how galaxies of different classes are affected by the cluster environment.}
{We applied the ROGER code to select trajectories of galaxies in the phase space for 35 galaxy clusters from the OmegaWINGS survey. A new algorithm was applied to minimize contamination effects.}
{We found that both morphological transformation and the quenching of star formation begin shortly after galaxies enter the cluster. Even though over the last $2-3$ Gyr, galaxies entering clusters have undergone significant transformations in both their star formation and morphology these transformation processes are not complete, that is, they are not completely quenched and are not early type yet. Backsplash galaxies and recent infallers show a higher fraction of jellyfish galaxies compared to older cluster members, suggesting that the timescale of this phenomenon is typically less than 3 Gyr.}{}

\keywords{
             galaxies: general --
             galaxies: stellar content --
             galaxies: evolution --
             galaxies: clusters: general
                             }

\maketitle


\section{Introduction}

Galaxy clusters represent the most densely populated virialized environments. They are defined by a strong gravitational potential, hot ionized intracluster gas, and may host up to thousands of galaxies. It is well established that various galaxy properties change systematically depending on their environment: morphology (e.g., \citealt{dressler80,Whitmore:1993,Dominguez:2001,Bamford:2009, Paulino-Afonso:2019, Vulcani23}), color (e.g., \citealt{Blanton05,martinez06,Weinmann:2006,martinez08}), 
luminosity (e.g., \citealt{Adami:1998,Coenda:2006}), the fraction of star-forming galaxies (e.g., \citealt{Hashimoto98,Mateus:2004,BlantonMoustakas:2009,Schaefer:2017,Coenda:2019, Vulcani10, Paccagnella16, Perez23}) and the gas content (e.g., \citealt{Giovanelli85, Cortese11, Brown17}). In cluster environments, galaxies generally evolve to become redder, to have earlier-type morphologies, and exhibit older stellar populations. Additionally, cluster galaxies show reduced gas content, lower levels of star formation, and weaker emission lines in their spectra than galaxies in less dense environments like in the field. 

Several mechanisms are responsible for influencing galaxies within clusters, particularly by depleting gas and subsequently halting star formation. One key process is ram-pressure stripping (RPS, e.g., \citealt{GG:1972}; \citealt*{Abadi:1999, Book:2010, Vijayaraghavan:2015, Steinhauser:2016}), which removes cold and warm gas from galaxies due to the pressure exerted by the intracluster hot gas. Another mechanism that can deplete the gas supply is tidal stripping caused by the gravitational potential of the cluster (e.g., \citealt{Gnedin:2003a, Villalobos:2014, Lopez2022}). 

In intermediate-density regions, such as cluster outskirts and groups, galaxy-galaxy interactions, also known as harassment, play a more significant role (e.g., \citealt{Spitzer51,Moore:1998}). This process can lead to both gas loss and morphological changes. 
Tidal stripping from these galaxy encounters can strip stars or truncate the stellar disc of galaxies, transforming them into spheroid-dominated systems (e.g., \citealt{Smith:2015}). 
Morphological evolution is thought to be largely driven by mergers (e.g., \citealt{Toomre:1977, Barnes:1992, Martin:2018}), which are common in galaxy groups but much rarer in clusters due to the high relative velocities of galaxies. Major mergers tend to form spheroidal systems \citep{Navarro:1994}, while gas-rich minor mergers may lead to the formation of massive disks \citep{Jackson:2022}.

Although the environment is important in shaping the properties of galaxies, internal processes that depend on stellar mass have proven to be very efficient in shutting down the star formation. Some of these processes are: supernova-driven winds (e.g., \citealt{Bower:2012, Stringer:2012}), halo heating \citep{Marasco:2012}, feedback from massive stars (e.g., \citealt{DallaV:2008, Hopkins:2012}), and active galactic nuclei (AGN) feedback (e.g.,\citealt{Nandra:2007, Hasinger:2008, Silverman:2008, Cimatti:2013}). These mechanisms are collectively known as ``mass quenching''.

It is well known that the observed properties of galaxies in clusters depend on the projected distance to the cluster center, with morphology being one of the first examples (\citealt{Whitmore91}). However, the projected distance is too imprecise to determine both the fall time and the type of orbit. Typically, the orbits of substructures falling into a cluster-size halos are highly radial and eccentric (\citealt{Gill05}). As a consequence, at a given radius within a cluster, there can be both newly infalling objects and others that have been orbiting for several gigayears. However, when considering both the cluster-centric distance and velocity with respect to the mean velocity of the cluster galaxies (i.e., the phase-space diagram), recent infallers generally exhibit higher velocities compared to long-term galaxies. This results in different regions of the phase space being more populated by objects with similar orbits and infall times. Moreover,  radial and eccentric orbits may result in the galaxy having a trajectory that crosses the virial radius after passing through the cluster's pericenter, which in phase space tends to place it at large projected radii and low velocities. These galaxies are known as backsplash \citep{Balogh00, Mamon04}.

\citet{Rhee17} correlate the position in the projected phase space with the time since infall (the time since a galaxy crossed for the first time the virial radius) and derive regions of constant mean infall time for galaxies in the different regions of the phase space. These authors divide the phase space into different zones that they associate with different infall times. Similar analyses have been conducted by \citet{Pasquali19}.

\citet{delosrios:2021} and \citet{Coenda:2022} proposed a novel method to separate galaxies in the phase space, taking into account the type of orbit. The technique employs machine learning algorithms to assign probabilities of having a specific type of orbit or the time spent within the cluster. To implement this algorithm, these authors developed the code ROGER (Reconstructing Orbits of Galaxies in Extreme Regions), which is publicly available~\footnote{There are two implementations available: in \texttt{R} \href{https://github.com/Martindelosrios/ROGER}{https://github.com/Martindelosrios/ROGER} and in python \href{https://github.com/Martindelosrios/pyROGER}{https://github.com/Martindelosrios/pyROGER}.} and will be described in section \ref{sec:ROGER}.

ROGER has been recently applied to a sample of X-ray-emitting clusters by \citet{Martinez23}. They observe that significant morphological evolution occurs only in virialized galaxies within clusters. However, they also note that blue galaxies recently entering the cluster may have already experienced morphological changes. They explore whether quenching timescales are generally shorter than those required for morphological changes. Their findings indicate that quenching happens more rapidly across all predicted dynamical categories. Moreover, while quenching is noticeably accelerated as soon as galaxies enter clusters, morphological transformations take longer times, requiring prolonged exposure to the cluster's physical mechanisms. In contrast, \citet{Sampaio24} analyzed a sample of clusters in the SDSS and found that morphological transition precedes complete star formation quenching.

\citet{Martinez23} also analyzed the galaxy populations by color for each of the classes provided by ROGER that are: cluster galaxies, backsplash galaxies, recent infallers, infallers, and interlopers. For cluster and backsplash galaxies, they limited the analysis to red galaxies due to the contamination produced by projection effects (see \citealt{Coenda:2022} for a detailed discussion). \citet{Martinez23} found that 
blue recent infallers tend to be smaller than both infalling galaxies and interlopers, which are similar in size. Based on these results, they suggested that a single pass through the cluster environment can alter the galaxy’s morphology, and also shrink the size of blue galaxies. \citet{Marasco23} analyze a sample of galaxies with MUSE data  in groups and clusters to explore how the aging of stellar populations can lead to a morphological change in galaxies where star formation  has been rapidly quenched. They found that the morphological transformation is completed after 1.5–3.5 Gyr, occurring faster for more efficient quenching scenarios. 

Galaxies that halt star formation in less than $1.5\,\mathrm{Gyr}$ exhibit distinctive spectral features, such as the absence of emission lines and $H_{\delta}$ absorption, and are referred to as post-starburst galaxies (PSB, \citealt{Dressler82}). Analyzing the characteristics of their stellar populations and their distribution within clusters can provide important insights into the physical mechanisms behind the suppression of star formation. Analyzing WINGS and OmegaWINGS cluster galaxies in phase space, \citet{Paccagnella17} found that PSBs comprise a mix of galaxies with diverse accretion histories. PSBs with the most pronounced $H_{\delta}$ feature seem to have been recently accreted. In terms of stellar masses, magnitudes, colors, and morphologies, they found that PSBs fall between passive and emission-line galaxies, characteristic of a population transitioning from star formation activity to a passive state.  Analyzing different environments, \citet{Paccagnella19} found that the fraction of PSBs increases with halo mass, suggesting that mechanisms common in densely populated regions, such as ram-pressure stripping, play a significant role in the formation of many PSB galaxies in these environments (see also \citealt{Socolovsky19,Vulcani20,Wilkingson21,Werle22}).

Ram-pressure stripping can effectively remove gas from galaxies during their initial entry into clusters, with galaxies on radial orbits experiencing more severe gas loss (e.g., \citealt{Jaffe:2015,Yoon:2017,Jaffe:2018}). Additionally, radio observations reveal disrupted and stripped gas while the galaxy's stellar core remains largely intact, a phenomenon consistent with RPS. In some cases, star formation has been detected in the stripped tails (e.g., \citealt{Gavazzi95,Poggianti19,Lee22,Roberts22,Gullieuszik23}). Due to the appearance caused by this effect, these galaxies are known as jellyfish galaxies. \citet{Jaffe:2018} studied the phase-space distribution of cluster galaxies and found that jellyfish galaxies tend to have higher peculiar velocities \citep{Biviano24} and more radially elongated orbits compared to the overall cluster population. They also found that those jellyfish with the longest gas tails are moving at high speeds, and close to cluster cores, where ram pressure is stronger. They concluded that many jellyfish galaxies likely formed through rapid, outside-in ram-pressure stripping during their first infall into the cluster. \citet{Salinas24} estimated the duration of optical tails of jellyfish galaxies, from their initial appearance to their eventual disappearance. They found that galaxy tails first emerge around $\sim$ 1.16 $R_{200}$ and vanish approximately 0.6 Gyr after reaching the pericenter.

In this work, we investigate the properties of galaxies in clusters depending on their orbit type using the predictions of the ROGER code. We are particularly interested in comparing galaxies that have been within the cluster from more than 2 Gyr ago and have completed several orbits (hereafter ``cluster galaxies''), with those that have made a single passage through the cluster center and are now located outside the cluster virial region (hereafter ``backsplash galaxies''). This comparison aims to understand the initial impact that the cluster environment exerts on galaxies during their first encounter. To achieve this goal, we analyze the morphology and age of the populations, as well as the fractions of passive galaxies, recently quenched galaxies, PSB, and jellyfish galaxies.

In this paper, we use a subsample of OmegaWINGS clusters (OW), which are described in section \ref{sec:Data}. The ROGER code is detailed in section \ref{sec:ROGER}, while in subsection \ref{sec:misclassification}, we present the technique used to correct for misclassification effects. The properties of the galaxies are analyzed in section \ref{sec:results}, where morphology, the fraction of quenched galaxies, age, and the fractions of PSB and jellyfish galaxies are compared by orbit. In section \ref{sec:Conclu}, we summarize our findings and present our conclusions. Throughout this paper we assume a concordance $\Lambda$CDM cosmology with $\Omega_M=0.3$, $\Omega_{\Lambda}=0.7$, and $H_0=70\ \mathrm{km\, s^{-1}\, Mpc}^{-1}$.


\section{Data}
\label{sec:Data}

\subsection{The cluster sample} The galaxy clusters used in this study constitute a subset derived from the OW spectroscopic survey (\citealt{Gullieuszik15}; \citealt{Moretti17}). This survey serves as an extension of the WIde-field Nearby Galaxy Cluster Survey (WINGS, \citealt{Fasano06}; \citealt{Varela09}; \citealt{Moretti14}), which is a multi-wavelength survey aimed to cover the outskirts of 76 massive clusters (of which, 46 were spectroscopically observed), spanning a redshift range of $0.04 \leq z \leq 0.07$, selected from the ROSAT All-Sky Survey \citep{Ebeling96}.

OW expands the spatial coverage of the WINGS survey for 46 of these clusters, acquiring imaging data in the $B$, and $V$ bands across an area of approximately $1\ \deg^2$. Moreover, it encompasses a broad range of velocity dispersion ($\sigma \sim 500-1,300\ \mathrm{km\, s}^{-1}$), X-ray luminosity ($L_X \sim 0.2-5\, \times 10^{44}\ \mathrm{erg\, s}^{-1}$), and $R_{200}$ (radius where the mean interior density is 200 times the critical density of the Universe) of about $1-3\, \mathrm{Mpc}$. The $R_{200}$ values for the cluster sample are taken from \cite{Biviano17}. The target selection process for the spectroscopic observations mirrored that of the photometric survey (\citealt{Cava09}; \citealt{Moretti17}). These selections were based on galaxies with a total magnitude brighter than $V = 20$, excluding those significantly above the color-magnitude sequence with $B - V > 1.20$. Additionally, galaxies brighter than $M_V = -17.4$ (the absolute magnitude limit of a galaxy with $V = 20$ at the redshift of the most distant cluster in the sample, reaching a magnitude completeness of $80\%$), were chosen (see \citealt{Paccagnella17}).
Spectroscopic follow-up for OW included observations using the VST fibre spectrograph. OW spectra cover a range of approximately $3800$ to $9000$ \AA, with a resolution of $3.5-6$ \AA. In this work, we have restricted the sample to clusters that have at least 50\% completeness in spectroscopy, resulting in a total of 35 clusters.

\subsection{The spectrophotometric code}
Although the sample of clusters and galaxies selected in this work is the same one used in \citet{Salerno20}, the spectroscopic properties of galaxies were recalculated following \citet{Fritz17}. The stellar population properties are determined using the spectrophotometric code SINOPSIS\footnote{\url{https://jacopofritz.wixsite.com/webpage/sinopsis}} (SImulatiNg OPtical Spectra wIth Stellar population models). SINOPSIS utilizes the theoretical spectra of simple stellar populations (SSP) across 12 different ages, ranging from $10^6\ \mathrm{yr}$ to the age of the Universe at the redshift of the galaxy ($t_u$); and four metallicity values: sub-solar (Z = 0.004), solar (Z = 0.017), and super-solar (Z = 0.03, 0.04).  The SSP models are based on work by Charlot \& Bruzual (in prep.), using a \citet{Chabrier03} IMF with masses ranging from 0.1 to 100 $M_\odot\ $. After obtaining the best fit, an age binning is applied, reducing the final resolution to four age bins. 

The stellar age bins are defined according to stellar population features \citep{Fritz07}, thus:
\begin{itemize}
    \item[$\star$] Stellar age bin 1 (0 - 19.95 Myr): Characterized by emission lines and the strongest ultraviolet emission.
    \item[$\star$] Stellar age bin 2 (19.95 - 571.5 Myr): Hydrogen lines from the Balmer series reach their maximum intensity in absorption, while the Ca{\sc k,h} UV lines still have low (almost undetectable) equivalent widths.
    \item[$\star$] Stellar age bin 3 (0.5715 - 5.754 Gyr): Balmer absorption lines decrease in intensity, while the \textsc{k} calcium line reaches its maximum level in absorption.
    \item[$\star$] Stellar age bin 4 (5.754 Gyr - $t_u$): SSPs in this age bin are reddest, and the main spectral characteristics show an asymptotic behavior. The $4000$ \AA \ break (D4000) attains the highest values.    
\end{itemize}
The resulting star formation rates (SFRs) for these four bins form the final star formation history (SFH). Total stellar mass is obtained by rescaling the aperture mass using the aperture and total magnitudes in the V-band. In this way, we are assuming that the color gradient between the aperture and the whole galaxy is negligible, as done by other authors (e.g., \citealt{Kauffmann03})
Detailed descriptions of the code can be found in \citet{Fritz07}, \citet{Fritz17}, and \citet{Perez23}.

To address incompleteness corrections (both geometrical and magnitude-based) in the spectroscopic catalogs, the ratio of spectra yielding a redshift to the total number of galaxies in the parent photometric catalog was utilized, computed as a function of $V$ magnitude and radial distance from the brightest cluster galaxy. All calculations were conducted while weighting each galaxy to account for both types of incompleteness. The ultimate spectroscopic sample comprises 14,801 galaxies within 35 out of 46 OW clusters.
 
In this study, we analyze galaxy properties based on morphological types, using the morphologies estimated by \citet{Fasano12} and \citet{Vulcani23}. These authors employ the MORPHOT code, a non-parametric and fully empirical method for automatically estimating galaxy morphology. MORPHOT utilizes 21 morphological diagnostics that can be directly computed from galaxy images and provides two distinct classifications: one derived from a maximum likelihood semi-analytical method and the other from a neural network approach. The final morphological estimator combines both techniques and has been tested on an additional sample of visually classified WINGS galaxies (see also, \citealt{Vulcani11b}).

The MORPHOT code assigns each galaxy a TYFIN type ranging from $-6$ to $11$. We classify them into two groups: early-types, which include galaxies with TYFIN $\leq$ 0 such as cD, elliptical, and S0 galaxies; and late-types, with TYFIN $>$ 0, corresponding to early spirals, late spirals, and irregular galaxies. 


\section{ROGER classification}
\label{sec:ROGER}

\subsection{Orbital classes}
Given the position of a galaxy in the Projected Phase Space Diagram (PPSD), specifically, its projected distance to the cluster center in terms of $R_{200}$, and its line-of-sight velocity relative to the cluster in units of $\sigma$, ROGER\footnote{In this work we used the python implementation of ROGER that is available in: \url{https://github.com/Martindelosrios/pyROGER}} was trained by \citet{delosrios:2021} to compute the probabilities for a galaxy to belong to different classes using three Machine Learning techniques: K-Nearest Neighbors (KNN), Support Vector Machines, and Random Forest. It is worth to mention that these machine learning methods were trained using galaxies from the MultiDark Planck 2  simulation \citep{klypin_mdpl2_2016}, where the real classes and properties of each galaxy are known. ROGER was trained using a sample of 34 massive clusters. 
This sample is more massive than that of OmegaWINGS (medians of $\log(M_{200})$ are 15.04 and 14.67, respectively) and have a larger velocity dispersion (median values of 1255 and 744 $ {\rm km/s}$, respectively). However, ROGER uses $\Delta v_{los}/\sigma$, which exhibits similar behaviors in both samples. This similarity is directly related to the self-similarity of halos, so the application of ROGER to a less massive sample does not introduce significant biases. ROGER computes, for each galaxy, its probability to belong to each of the following five orbital classes, as defined by \citet{delosrios:2021}:
 
(1) Cluster galaxies (CL): Galaxies designated as satellites of the cluster, having maintained this status for over 2 Gyr. The majority reside within $R_{200}$, with a few exceptions ($\sim 4$\%), temporarily beyond $R_{200}$ in their orbital trajectories.
 
(2) Backsplash galaxies (BS): Galaxies that have traversed $R_{200}$ precisely twice, once on their entry into the cluster and again on their exit. Located outside $R_{200}$, these galaxies have undergone a single transit through the cluster core. They are likely to eventually become CL in the future.
 
(3) Recent infallers (RIN): These are galaxies located within $R_{200}$, having crossed it just once on their entry and no earlier than 2 Gyr ago. These galaxies are currently undergoing the environmental effects of the cluster for the first time. Some may transition into backsplash galaxies (BS) in the future.
 
(4) Infallers (IN): These galaxies have spent their entire lifetimes outside $R_{200}$ and are currently in the process of falling into the cluster, indicated by their negative radial velocities relative to the cluster.
 
(5) Interlopers (ITL): These galaxies have likewise remained outside $R_{200}$ for the entirety of their lifetimes; however, they are not approaching the cluster and lack any physical association with it. Their presence in the PPSD is solely attributable to projection effects.

Like any selection method based on the phase space, the selection of the orbital classes by ROGER is affected by potential deviations from the state of relaxation. Since ROGER is trained with galaxies in dark matter halos selected from the MultiDark cosmological simulations, it exhibits the same biases as an observed sample. In the present analysis, we will consider categories CL, BS, RIN, and ITL. The IN class was not included in our analysis, as it does not represent a statistically significant sample. Although $\sim$12\% of the total sample has the highest probability corresponding to the IN class, this percentage is reduced to less than 2\% (91 galaxies across the entire stellar mass range) when applying the \citet{Coenda:2022} criteria\footnote{In order to reduce contamination among classes in the PPSD diagram, \citet{Coenda:2022} tested various threshold values and selected those that produce pure samples (high precision) for each class without sacrificing statistical power (high sensitivity). They proposed the optimal threshold values for each class.}, suggesting that the sample of galaxies classified as IN based on the highest probability could be heavily contaminated by other classes. On the other hand, galaxies classified as ITL should not necessarily be interpreted as field galaxies, as due to hierarchical clustering, a significant fraction of these galaxies could belong to groups or filaments around the clusters.

\begin{figure}
\centering
\includegraphics[width=.85\columnwidth]{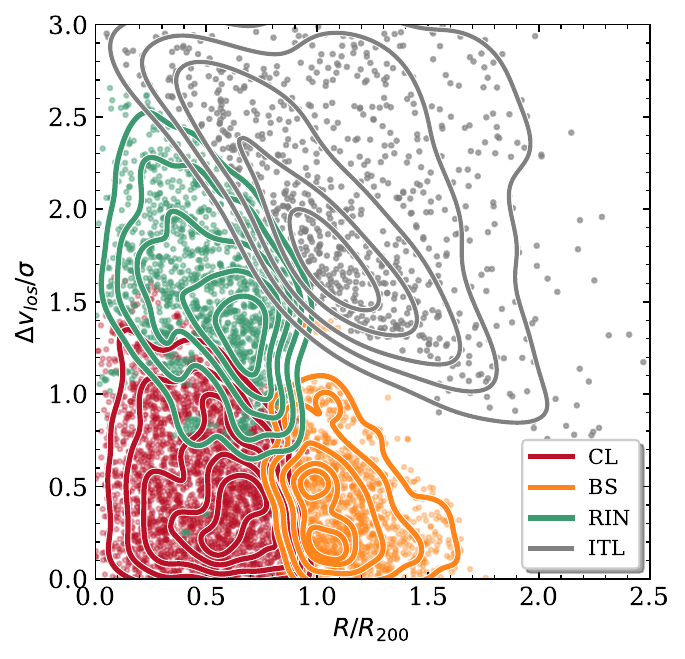}
\caption{Distribution of galaxies in the phase space according to the classification produced by ROGER. The types of orbits considered are: cluster galaxies (CL, red); backsplash galaxies (BS, orange); recent infallers (RIN, green) and interlopers (ITL, grey).
}
\label{fig:fase}
\end{figure}

Based on the findings of \citet{delosrios:2021}, we employ the trained KNN technique to determine class probabilities for our galaxies. 
ROGER was applied to our data and each galaxy was assigned an orbit type based on its highest probability of belonging to a given orbit and weighted by the probability to belong to this category. The resulting number of galaxies in each predicted class is provided in Table \ref{tab:numbers}.
In Fig. \ref{fig:fase} we display the PPSD positions of galaxies that meet our classification criteria, represented by different colors.
 
\begin{table}
\caption{Total number of galaxies and by morphological type for each of the types of orbits considered. }
\center
\small
\begin{tabular}{rcccccccc}
\hline \hline
\!\!\!Property & CL & BS & RIN & ITL \\
\hline
      All   & 2080 (669) & 997 (393)   & 928 (550) &  530 (477) ~ \\
      ET    & 1424 (484) & 524 (198) & 587 (339) & 222 (187) \\
      LT    & 524 (147) & 314 (132) & 288 (180) & 238 (230) \\
\hline \hline
\end{tabular}
\tablefoot{The numbers in parentheses correspond to the weighted values (see text for more details).}
\label{tab:numbers}
\end{table}


\subsection{Correcting the effects of misclassification}
\label{sec:misclassification}

The determination of orbits in the phase space is strongly limited by issues of contamination and incompleteness, we refer the reader to \citet{Coenda:2022} for a thorough  discussion on this matter. To mitigate these problems, \citet{Martinez24} proposed a method to improve the accuracy of determining the distribution of galaxy properties by statistically correcting for misclassification errors in the PPSD. The method involves the use of an estimation the confusion matrix, which contains the information of how each class is contaminated from misclassified galaxies that actually belong to the other classes. By inverting this matrix, the authors are able to obtain better determinations of the intrinsic distributions of galaxy properties, such as color. The method was tested using simulated galaxy clusters, showing significant improvements in the accuracy of classifying galaxies, particularly for those in cluster and backsplash categories. The corrected distributions eliminate most of the contamination from misclassified galaxies, offering a more reliable analysis of galaxy properties in various environments. The authors emphasize that their method allows for a more precise interpretation of observational data, reducing biases caused by misclassifications.

The method of \citet{Martinez24} allows us to compute for each galaxy in our samples an individual weight depending on its class $i$, mass $M$, and rest-frame $B-V$ color. We proceed as follows:
\begin{enumerate}
    \item We split galaxies into four bins in $\log(M/M_{\odot})$: 
    $[9.48\footnote{According to \citealt{Perez23}, the sample is complete in stellar mass for $\log(M/M_{\odot}) \geq 9.48$.},9.8)$, $[9.8,10.2)$, $[10.2,10.6)$, and $[10.6,11.5]$.
    \item For each class $i$ and mass bin $j$ we compute the observed color distribution using 11 equal size
    bins in $B-V$ color, ranging from $-0.3$ to $1.7$: $f_\mathrm{obs}(i,j,k)$, where $k$ denotes the bin in color.
    \item Following \citet{Martinez24}, we obtain, for each class and mass range, the reconstructed 
    color distribution $f_\mathrm{rec}(i,j,k)$. We use the confusion matrix estimated by the authors.
    \item The weight we assign to a galaxy of class $i$, stellar mass in the $j-$th bin, and color in the $k-$th 
    bin is the quotient: $f_\mathrm{rec}(i,j,k)/f_\mathrm{obs}(i,j,k)$.
\end{enumerate}
 The derived weights are $\leq$ 1 and have a median of approximately 0.6.


\section{Results: Comparing physical properties of CL, BS, RIN, and ITL galaxies}
\label{sec:results}

\subsection{Morphology}

\begin{figure}
\centering
\includegraphics[width=.9\columnwidth]{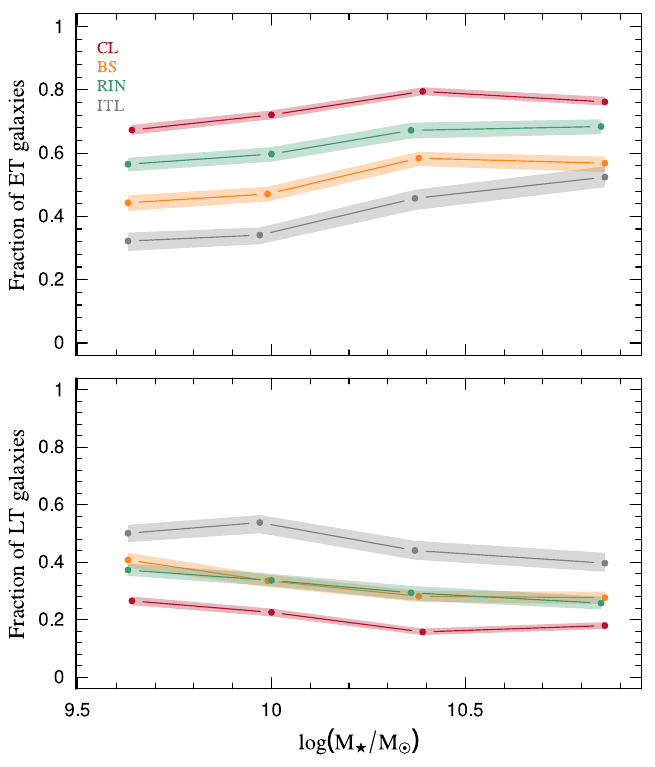}
\caption{Fractions of early (\textit{upper panel}) and late (\textit{bottom panel}) type galaxies as a function of the stellar mass for CL, BS, RIN, and ITL galaxies.}
\label{fig:morph}
\end{figure}

As discussed in the introduction, a single passage through the inner regions of the cluster can not only alter the star formation  but also modifies the morphology of a galaxy. Depending on the ROGER class, the morphological type has been estimated for 80 to 90\% of the galaxies. Fig. \ref{fig:morph} shows fractions of early and late type galaxies (ET and LT, respectively) as a function of stellar mass for CL, BS, RIN, and ITL classes. In this and the following figures, errors were obtained using the bootstrap resampling technique. 
The fractions of early- and late-type are not mirror-like for two reasons: in addition to the fact that not all galaxies in the sample have been morphologically classified, each galaxy has an assigned weight depending on its characteristics.
The first thing we observe is that, regardless of the type of orbit, the ET fractions systematically increase (and LT fractions decrease) with stellar mass. Similar results were found by \cite{Vulcani11a} and \cite{Vulcani11b}. If we compare the fractions at fixed stellar mass bin, we observe that the highest fractions of ET correspond to cluster members and the lowest values to ITL. BS and RIN have intermediate values between the two previous ones, with RIN having higher fractions of ET than BS. For the LT fractions, we again find that CL and ITL are the extreme cases, with CL having the lowest LT fractions, which, depending on the stellar mass, ranges between 0.2 and 0.3. While BS and RIN also exhibit intermediate fractions, in this case, they are indistinguishable from each other.

\begin{figure}[h]
\centering
\includegraphics[width=1.\columnwidth]{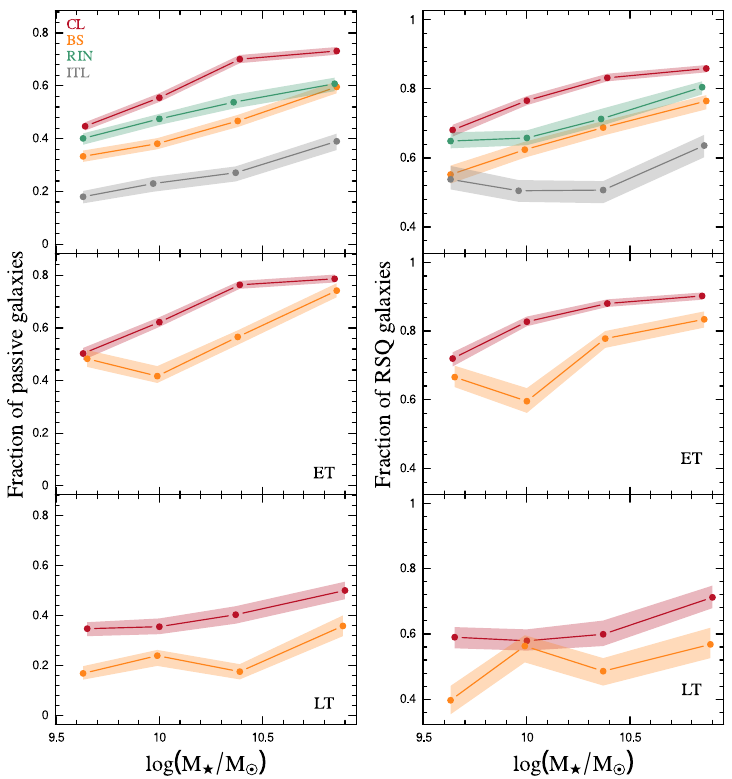}
\caption{Left panel: Fractions of CL, BS, RIN, and ITL passive galaxies  as a function of stellar mass. Top panel: All morphological types; central and bottom panels: Early and late types, respectively, for CL and BS galaxies. Right panel: Fraction of recently strongly quenched galaxies. Types of galaxies are presented as shown in the left panel.}
\label{fig:Frac_SFR_RSQ}
\end{figure}


\subsection{Star formation quenching}
The left column of Fig. \ref{fig:Frac_SFR_RSQ} shows the fractions of CL, BS, RIN and ITL passive galaxies in four bins of stellar mass (we assume that a galaxy is passive if its SFR~$\leq 0.0001\ M_{\odot}\, \mathrm{yr}^{-1}$). In agreement with what is observed for ET galaxy fractions (see Fig. \ref{fig:morph}), the highest (lowest) fraction of passive galaxies corresponds to CL (ITL), while BS and RIN have intermediate values. RIN have a higher fraction of passive galaxies than BS across almost the entire stellar mass range. As expected, in all cases, an increase in the fraction of passive galaxies is observed as stellar mass increases. 

For CL and BS galaxies, we have also analyzed the fractions of passive galaxies as a function of morphological types. Since CL and BS are mostly ET, they show similar behavior to the overall sample (see central left panel of Fig. \ref{fig:Frac_SFR_RSQ}), although with systematically higher values, while LT galaxies (lower panel) show a weaker dependence on stellar mass. This could be indicating that low mass LT galaxies are being more affected by the environment, compensating for the lower quenching they experience by the mass-quenching compared to more massive ones. These two opposing phenomena tend to flatten the dependence of star formation quenching on mass.

With the aim of assessing how intense the recent quenching of star formation has been, we have calculated the fraction of galaxies that have experienced strong star formation quenching over the last $\sim 0.5$ Gyr. This fraction, which we refer to as RSQ (galaxies that have been Recently Strongly Quenched), is calculated as the number of galaxies per class that have been completely quenched, or have reduced their star formation by half during the mentioned period, divided by the total number of galaxies in the same class that were star-forming at the beginning of the interval.  In practice, the calculation involves comparing the SFR between the age bins $0-19.95$ Myr and $19.95-571.5$ Myr (bin 2, SFR$_2$), which are the first two age bins computed by the SINOPSIS code (see Table 1 in \citealt{Perez23}). According to \citet{Ruiz23}, backsplash galaxies at $z=0$ have been outside $R_{200}$ for 1.25 Gyr on average, meaning that the period over which we calculate the fraction is less than half the time the galaxies have been in the backsplash phase. 
In other words, most of the BS galaxies should already be outside $R_{200}$ by the age of bin 2, while most of the RIN galaxies should have already entered the cluster in the same period. For this calculation, we consider star-forming galaxies to be those with SFR$_2$ $\geq 0.001 M_{\odot} \,\mathrm{yr^{-1}}$; however, the general conclusions do not depend on the choice of this threshold. The RSQ fraction have been computed as a function of stellar mass and are shown in the right panel of Fig. \ref{fig:Frac_SFR_RSQ}.

The behavior of the RSQ fractions is very similar to that of the fraction of galaxies with SFR $=0$; in other words, the galaxies that have experienced the strongest quenching are the cluster galaxies. It is important to note that the calculation of the RSQ fraction is based on the sample of galaxies that were star-forming in age bin 2, which, in the case of CL galaxies, are the fewest. Of the small number of CL galaxies that were star-forming in bin 2, approximately 80\% have been strongly quenched, whereas this percentage is 50\% for ITL. RIN galaxies, despite their short time in the cluster, show a significant degree of quenching, although clearly less than that of the CL galaxies. The BS galaxies, although they were mostly already outside $R_{200}$ in age bin 2, have continued their quenching process, something noted by \citet{Ruiz23}. 


\subsection{Age}

\begin{figure}
\centering
\includegraphics[width=.75\columnwidth]{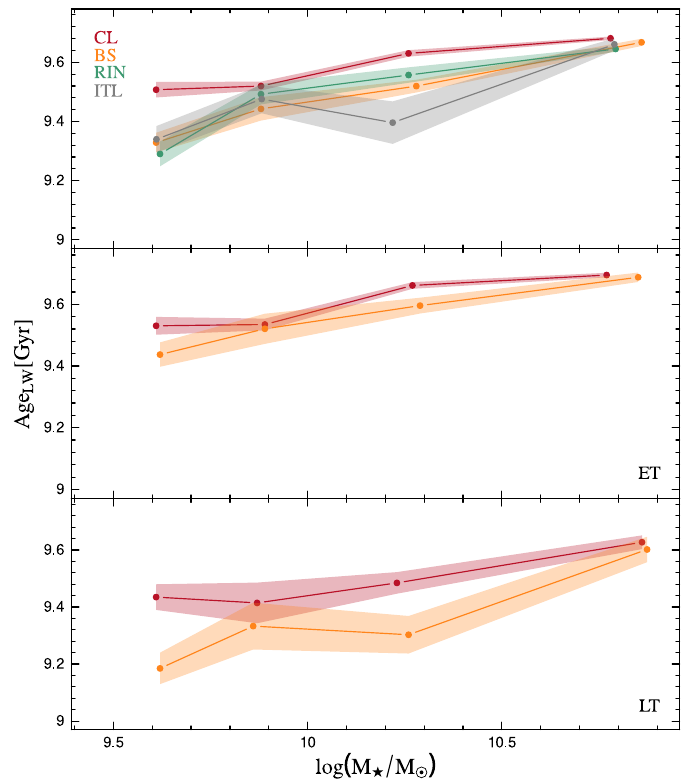}
\caption{Luminosity-weighted Age as a function of the stellar mass. Types of galaxies are presented as shown in Fig. \ref{fig:Frac_SFR_RSQ}.}
\label{fig:Age}
\end{figure}

Galaxies with different dynamical histories are expected to show differences in luminosity-weighted age. The top panel of Fig. \ref{fig:Age} shows that CL are on average always the oldest at all masses. When comparing the ages of BS, RIN, and ITL galaxies, no statistically significant difference is observed between these populations, which could be related to the fact that BS and RIN galaxies have spent most of their lives outside the cluster. Although these populations have shown clear differences in the degree of star formation quenching, this has not been sufficient to alter the average age of the samples.

The middle and lower panels of Fig. \ref{fig:Age} compare the ages of CL and BS galaxies for ET and LT galaxies, respectively, showing that the differences observed for the total samples are mainly due to late-type galaxies.

\subsection{Post-starburst galaxies}

Any process of star formation quenching can be studied by analyzing the population of galaxies that, although no longer forming stars, have done so recently. This type of galaxy is known as post-starburst galaxies, and they can be identified through spectral analysis. \citet{Fritz2014} identified in the WINGS survey k+a/a+k spectra that exhibit a combination of features characteristic of both K and A-type stars, with strong $H\delta$ absorption lines and an absence of emission lines. These are typical of post-starburst/post-starforming galaxies where star formation was abruptly halted at some point within the last 0.5–1 Gyr.

\begin{figure}
\centering
\includegraphics[width=.7\columnwidth]{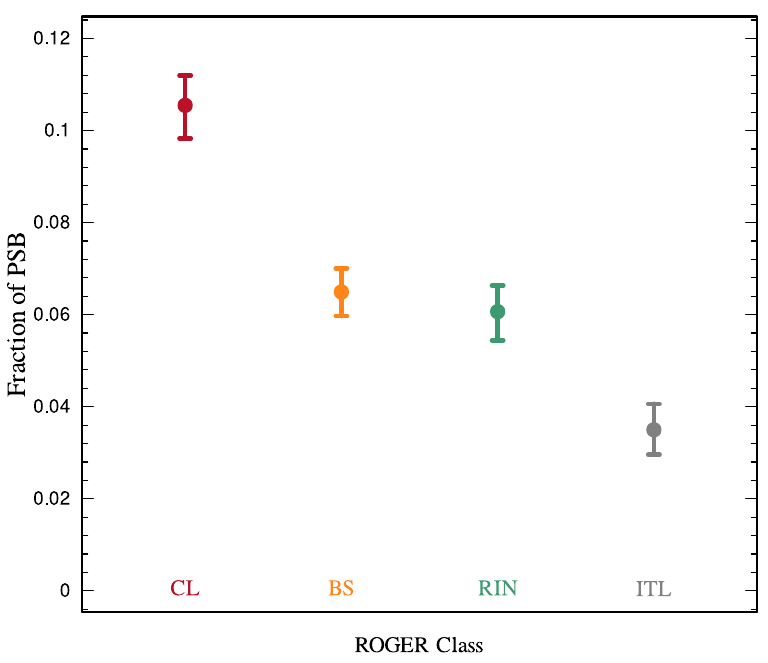}
\caption{Fractions of post-starburst galaxies within the predicted classes CL, BS, RIN, and ITL.} 
\label{fig:PSB}
\end{figure}

Using the criteria established by \citet{Fritz2014} to classify PSB galaxies, \citet{Paccagnella17} and \citet{Paccagnella19} identified PSB galaxies in OmegaWINGS cluster environments. Fig. \ref{fig:PSB} shows the PSB fractions for CL, BS, RIN, and ITL galaxies. Although the fractions are very small in all cases, it can be seen that cluster galaxies have the highest fraction of PSB, while ITL show the lowest fraction. With intermediate values, BS and RIN galaxies do not show statistically significant differences between each other. The fact that CL galaxies have the highest fraction of PSB, aligns with what was observed in the RSQ fractions\footnote{Note that the PSB fraction refers to all galaxies, while the RSQ fraction refers only to star-forming galaxies in age bin 2.}, which indicated that clusters have  remained efficient in quenching the star formation.

\subsection{Jellyfish galaxies}

The quenching of star formation is directly related to the loss of gas in galaxies. One of the most extreme examples of gas loss are the so-called Jellyfish galaxies, where, due to the action of the intracluster medium gas, galaxies can lose high percentages of their gas. This phenomenon can be further intensified by tidal forces, either from the cluster potential or through interactions with other galaxies. 

The pioneering work of \cite{GG:1972} showed that both the velocity of galaxies and the density of the intracluster medium are fundamental parameters in gas removal. The presence of Jellyfish galaxies will also depend on various factors, such as the time spent in the cluster, how close to the cluster center a galaxy passes, the duration of the Jellyfish phenomenon, the amount of gas available to be removed and the inclination of the galaxy relative to the velocity with respect to the intracluster medium (e.g., \citealt{Kronberger08}). Since the different orbit classes provided by ROGER occupy different regions in phase space, the fraction of Jellyfish galaxies is also expected to depend on the orbit type.

\citet{Poggianti16} have conducted the first systematic search for galaxies that are being stripped of their gas at low redshift, selecting galaxies with different degrees of morphological evidence for gas stripping. They have visually inspected images and identified 344 candidates in 71 galaxy clusters of the  OW+WINGS sample. The authors tentatively categorized candidates into five groups based on the visual signs of stripping observed in the optical bands (JClass), ranging from the most extreme cases (JClass$\,=5$) to progressively less intense ones, with JClass$\,=1$ being the least pronounced. Consequently, the JClass$\,=5$ and 4 categories include the most reliable candidates and feature the most notable examples of classic jellyfish galaxies. JClass$\,=3$ candidates are likely undergoing stripping and/or experiencing ram pressure events, whereas JClasses$\, =2$ and 1 are tentative cases for which the current imaging does not allow for definitive conclusions. \citep{Vulcani22} identified an additional subsample of ram pressure stripping candidates that were missed by the visual inspection performed by \citep{Poggianti16}.

\begin{figure}
\centering
\includegraphics[width=1.\columnwidth]{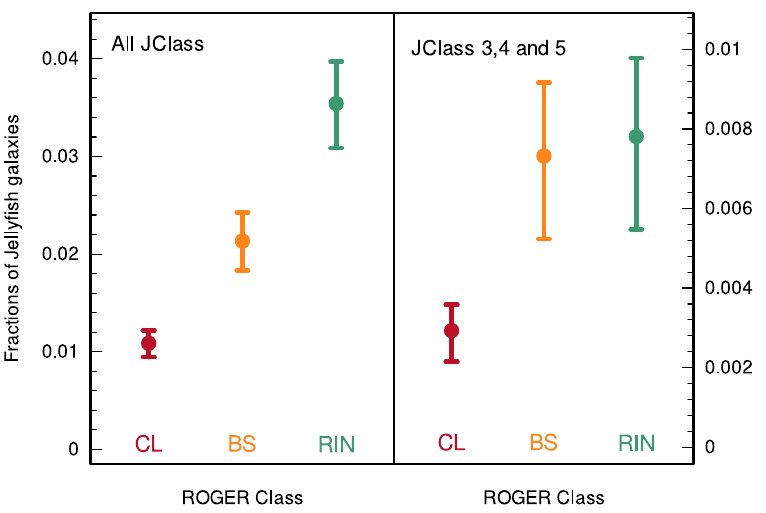}
\caption{Fraction of jellyfish galaxies within the predicted classes CL, BS, and RIN. b) As in panel a) for JClass 3, 4, and 5.}
\label{fig:JF}
\end{figure}

In the left panel of Fig. \ref{fig:JF}, the fractions of galaxies classified as jellyfish in any of the classes are shown. ITL galaxies were excluded from this analysis because, being objects external to the cluster, a significant fraction of the ITL-jellyfish is expected to be linked to galaxy interactions rather than effects caused by the intra-cluster gas. As expected, the fractions are extremely low, but a clear difference by class is observed, with CL showing the lowest fraction and RIN the highest, while BS have intermediate values. The right panel of Fig. \ref{fig:JF} shows the fraction of jellyfish galaxies for JClass values 3, 4, and 5, where we can see that CL galaxies continue to display the lowest fraction, while BS and RIN are indistinguishable, though with greater uncertainties compared to those presented in the left panel.


\section{Summary and conclusions}
\label{sec:Conclu}

We have studied different properties of cluster galaxies using the ROGER code, which allows the selection of the most probable orbits of galaxies in the projected phase space. For the first time in this type of studies, a technique was applied to minimize the consequences of misclassification caused by projection effects \citep{Martinez24}, which introduce serious biases in galaxy samples and, consequently, in the conclusions drawn from them (see \citealt{Coenda:2022}). The types of orbits analyzed are: cluster galaxies (CL), backsplash galaxies (BS), 
recent infallers (RIN), and interlopers (ITL). We focus on comparing cluster galaxies that likely orbited the cluster more than once, with backsplash galaxies that have made a single passage and are now outside the cluster. The main goal is to understand the initial impact of the cluster environment on galaxies after their first incursion within the virial radius.

For this analysis, a subsample of 35 OW clusters was used, and the following galaxy properties were considered: morphology, fraction of quenched galaxies, age, fraction of PSB galaxies, and fraction of jellyfish galaxies. In all cases, the analyses were performed as a function of the galaxies' stellar mass. Our main conclusions are:

\begin{enumerate}
\item Regardless of class, ET fractions increase and LT fractions decrease with stellar mass in agreement with previous results based on the same data (\citealt{Vulcani11a, Vulcani11b, Perez23}). Separating galaxies by class, higher ET fractions are found in cluster members, while ITL show the lowest values. BS and RIN have intermediate values, with RIN showing the highest ET fraction. For LT fractions, CL has the lowest values, ranging from 0.2 to 0.3 depending on stellar mass. BS and RIN show similar, indistinguishable LT fractions. The fact that the morphological type fractions of BS and RIN are clearly different from those of ITL suggests that morphological transformation begins shortly after galaxies enter the cluster, although it is clearly not yet complete, as clusters exhibit significantly higher ET fractions than those of BS and RIN. These results are consistent with those reported by \citealt{Martinez23, Vulcani11a, Vulcani11b}.
 
\item  BS and RIN galaxies exhibit significantly higher fractions of quenched galaxies compared to ITL, although lower than CL, indicating that the quenching of star formation proceeds rapidly once galaxies enter the cluster. Based on the fraction of RSQ (Recently Strongly Quenched) galaxies, we found that galaxies that have experienced the strongest quenching during the last 0.5 Gyr  are the cluster galaxies. Approximately 80\% of the CL galaxies have been strongly quenched during this period of time, whereas this percentage is 50\% for the ITL galaxies. It is important to clarify that the fractions of RSQ galaxies are calculated from galaxies that were still forming stars at $z\sim 0.04$, which, in the case of clusters, represent a smaller fraction compared to other orbital types. What our results indicate is that the few cluster galaxies still forming stars 0.5 Gyr ago, were rapidly quenched. This phenomenon is consistent with the slow-then-rapid quenching scenario (e.g., \citealt{Wetzel13, Maier19}), in which slow quenching, or ``strangulation'', begins when a galaxy crosses the $R_{200}$ radius and gas inflow is halted. During this phase, galaxies continue forming stars but show elevated metallicities as quenching begins. As they move toward the cluster’s denser core, a rapid phase follows in which star formation is abruptly halted due to increasing ram pressure, which can also strip away cold gas, even in massive galaxies.  RIN galaxies, despite their short time in the cluster, also show a significant degree of quenching. BS galaxies, although they were mostly already outside $R_{200}$ in age bin 2, have continued their quenching process, something noted by \citealt{Ruiz23}. 

\item  The mean ages of BS, RIN, and ITL galaxies are statistically indistinguishable and clearly younger than cluster members. Although these populations show clear differences in the degree of star formation quenching, this fact has not been sufficient to alter the average age of the samples. When analyzed by morphological type, the differences between BS and CL are mainly due to late-type galaxies.  

\item In agreement with \cite{Paccagnella17},  cluster galaxies have the highest fraction of PSB, while ITLs exhibit the lowest values. BS and RIN galaxies show intermediate values with no statistically significant differences between them. Although the fractions of PSB and RSQ galaxies have been calculated with respect to different sub-samples\footnote{The PSB fractions are calculated for all galaxies in a given class, while the RSQ fractions are based on those that were star forming 0.5 Gyr ago.
}, the fact that CL class (composed of galaxies that have been in the cluster for more than 2 Gyr) shows the highest fractions of both PSB and RSQ suggests that some galaxies require a longer time to be completely quenched. On the other hand, although BS and RIN galaxies exhibit significant fractions of quenched galaxies, there is still a considerable number of galaxies available to be quenched during successive orbits within the cluster in order to become PSB.

\item  The observed fractions of jellyfish galaxies clearly depend on the type of orbit, with CL showing the lowest fraction and RIN the highest, while BS have intermediate values. Considering that the jellyfish phase lasts between 1 and 2 Gyr (e.g., \citealt{Jaffe:2018}), it is expected that RIN, which entered the cluster less than 2 Gyr ago, are the most affected. On the other hand, BS galaxies, which according to \citet{Ruiz23} spend on average 1.8 Gyr in a cluster, would have already experienced the jellyfish phenomenon, and for some of them, the phenomenon might no longer be observable. Finally, CL galaxies, which have been inside the cluster for more than 2 Gyr, are not expected to be observed in their jellyfish phase. These results are in agreement with \citet{Yun19} who found that, typically, jellyfish galaxies are late infallers ($<2.5-3$ Gyr ago, at $z = 0$) in the TNG100 simulations. Similarly, \citet{Rohr23}, using the TNG50 cosmological simulation, found that the peak RPS period begins within approximately 1 Gyr of infall and lasts for $\sim 2$ Gyr.
\end{enumerate}

Our results suggest that, although over the past $2-3$ Gyr galaxies entering clusters have undergone significant transformations in both their star formation and morphology, it will still be necessary for them to orbit within the cluster for a longer time to complete the transformation processes.

Galaxies located outside $R_{200}$ that have already made their first passage through the cluster center continue undergoing transformations, particularly the quenching of the star formation. Considering that an average backsplash galaxy at $z=0$, according to \citet{Ruiz23}, spends $\sim$ 1.8 Gyr in its diving phase and $\sim$ 1.2 Gyr outside $R_{200}$, and that the fraction of passive BS galaxies is clearly lower than that in clusters, we conclude that the quenching process of a galaxy entering the cluster can last more than 3 Gyr. However, the fact that RIN, which on average should have spent less time in the cluster than BS, and which show a higher fraction of passive galaxies than BS, we cannot rule out the possibility that, despite the decontamination processes applied in this work, some degree of contamination between different orbital types still persists. In particular, BS may be experiencing some level of contamination from infalling galaxies and RIN from CL.  
An alternative explanation for the observed differences between BS and RIN arises when considering the times at which those two classes first dived into clusters. BS galaxies crossed the clusters' core when the clusters were less massive. 
At a lookback time of $\sim 3$ Gyr, clusters can be up to some $\sim 20\%$ less massive than at $z=0$ (see for instance \citealt{Vanderburg15}). Therefore, these galaxies were affected by the environment of typically lesser massive clusters than the one acting upon RIN at the present, and thus less affected by the environment during their passage through the clusters inner regions compared to RIN that are still moving inside the clusters. Furthermore, while BS galaxies were crossing the clusters, those that would later become RIN galaxies were being pre-processed by the environment in the outskirts of the clusters, adding up to the mass quenching experienced by all galaxies (note that BS and RIN galaxies are approximately the same age). 
This model is consistent with results recently found by Levis et al. (in prep.). These authors analyze a sample of groups of galaxies in the TNG simulations and find that the timing of a galaxy's entry into the group significantly influences its evolution, with later-arriving RIN galaxies experiencing stronger effects from the intragroup medium than earlier-arriving BS galaxies.


\begin{acknowledgements}
The authors thank the referee for her/his comments and suggestions that helped us improve the clarity of the paper. This paper has been partially supported with grants from Consejo Nacional de Investigaciones Cient\'ificas y T\'ecnicas (PIP 11220210100064CO), Argentina, the Agencia Nacional de Promoci\'on Cient\'ifica y Tecnol\'ogica (PICT 2020-3690, PICT-2021-I-A-00700), Argentina, Secretar\'ia de Ciencia y Tecnolog\'ia, Universidad Nacional de C\'ordoba, Argentina. MdlR is supported by the Next Generation EU program, in the context of the National Recovery and Resilience Plan, Investment PE1 – Project FAIR ``Future Artificial Intelligence Research’' and acknowledges financial support from the Comunidad Aut\'onoma de Madrid through the grant SI2/PBG/2020-00005.
D.P.M acknowledges financial support from the UNAM-DGAPA-PAPIIT IN111620 grant, Mexico, and from a CONAHCyT scholarship.
\end{acknowledgements}


\bibliographystyle{aa} 
\bibliography{references} 
 
\end{document}